\documentclass[aps,prl,twocolumn,superscriptaddress,showpacs]{revtex4}
\usepackage{graphicx}

\usepackage{dcolumn}

\usepackage{bm}
\usepackage{color}
\usepackage[colorlinks]{hyperref}
\input{epsf}

\begin{document}


\title{Demonstration of a 1/4 cycle phase shift in the radiation-induced oscillatory-magnetoresistance in GaAs/AlGaAs
devices}

\author{R. G. Mani}
\affiliation {Harvard University,
Gordon McKay Laboratory of Applied Science, 9 Oxford Street,
Cambridge, MA 02138, USA}

\author{J. H. Smet}
\author{K. von Klitzing}
\affiliation {Max-Planck-Institut f\"{u}r Festk\"{o}rperforschung,
Heisenbergstrasse 1, 70569 Stuttgart, Germany}

\author{V. Narayanamurti}
\affiliation {Harvard University, Gordon McKay Laboratory of
Applied Science, 9 Oxford Street, Cambridge, MA 02138, USA}
\affiliation {Harvard University, 217 Pierce Hall, 29 Oxford
Street, Cambridge, MA 02138, USA}

\author{W. B. Johnson}
\affiliation {Laboratory for Physical Sciences, University of
Maryland, College Park, MD 20740, USA}

\author{V. Umansky}
\affiliation{Braun Center for Submicron Research, Weizmann
Institute, Rehovot 76100, Israel}
%
%
%
%
\date{\today}
\begin{abstract}
We examine the phase and the period of the radiation-induced
oscillatory-magnetoresistance in GaAs/AlGaAs devices utilizing
in-situ magnetic field calibration by Electron Spin Resonance of
DiPhenyl-Picryl-Hydrazal (DPPH). The results confirm a
$f$-independent 1/4 cycle phase shift with respect to the $hf =
j\hbar\omega_{c}$ condition for $j \geq 1$, and they also suggest
a small ($\approx$ 2$\%$) reduction in the effective mass ratio,
$m^{*}/m$, with respect to the standard value for GaAs/AlGaAs
devices.
\end{abstract}
%
%
\maketitle

Measurements \cite{1} of GaAs/AlGaAs devices including a
2-Dimensional electron system (2DES) have recently shown that
vanishing resistance can be induced by photo-excitation at liquid
helium temperatures, in a weak-magnetic-field large-filling-factor
limit.\cite{1,2,3} The possibility of identifying a new physical
mechanism for inducing vanishing resistance in the 2DES has
motivated much theoretical interest in this
phenomenon,\cite{4,5,6,7,8,9} which shares features of the
zero-resistance states in the quantum Hall situation, minus Hall
quantization.\cite{10,11}

Experiments suggest that this novel effect is characterized, in
the Hall geometry, by a wide magnetic field, $B$, interval where
the diagonal resistance, $R_{xx}$, becomes exponentially small
upon photo-excitation in the low temperature, $T$,
limit.\cite{1,2,12} Measurements have also indicated that
increasing the radiation frequency, $f$, linearly shifts
distinguishing features to higher $B$.\cite{13,14} Thus, one might
define from $f$ a field scale $B_{f} = 2\pi f m^{*}/e$, where $e$
is the electron-charge, and $m^{*}$ is an effective mass.\cite{1}
According to theory, the observed periodicity reflects the
interplay between the photon energy, $h\emph{f}$, and integral
($j$) cyclotron energies, $j \hbar\omega_{c}$, and $B_{f}$ is
defined by $hf = \hbar \omega_{c}$.\cite{4,6,8} Hence, in the
expression for $B_{f}$, $e$ is a known constant, $f$ can be
measured accurately, and $m^{*}$ can be taken, initially, to be
the standard value for $m^{*}$ in GaAs/AlGaAs, although it could
involve corrections.
\begin{figure}
\begin{center}
\includegraphics*[scale = 0.25,angle=0,keepaspectratio=true,width=3.25in]{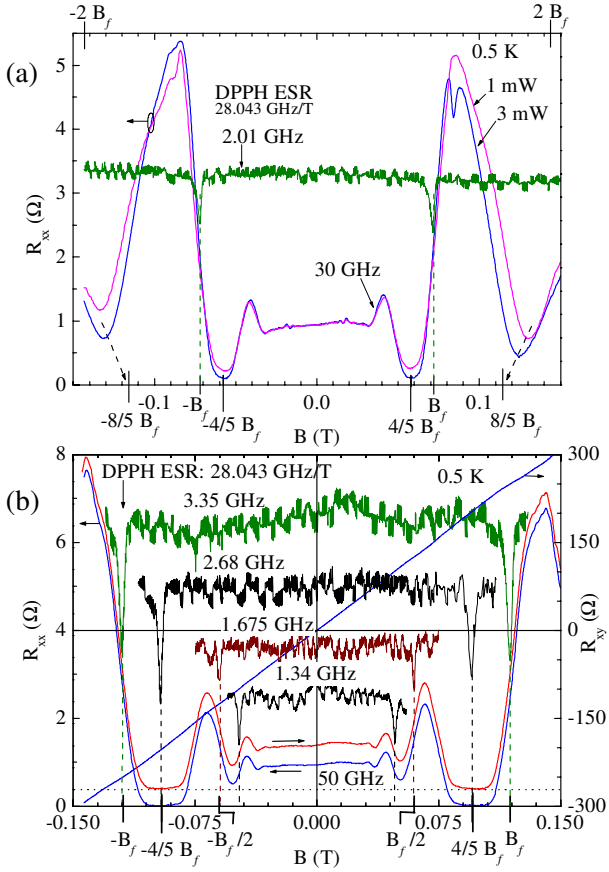}
\caption{(a) The magnetoresistance $R_{xx}$  is plotted vs. the
magnetic field $B$ for a GaAs/AlGaAs device under photoexcitation
at $f$ = 30 GHz. Also shown is the Electron Spin Resonance (ESR)
of DiPhenyl-Picryl-Hydrazal (DPPH) at 2.01 GHz, which marks
$B_{f}=2\pi f m^{*}/e$. $R_{xx}$ exhibits neither a maximum nor a
minimum at $B_{f}$. A deep $R_{xx}$ minimum between $B_{f} < B <
2B_{f}$ appears to converge to $2(4/5 B_{f}) = 8/5 B_{f}$,
suggestive of a two photon effect. (b) $R_{xx}$ and $R_{xy}$ at 50
GHz along with DPPH-ESR field markers for $B_{f}$, $4/5 B_{f}$,
$B_{f}/2$, and $B_{f}/2.5$. Note the similarity in the up- and
down- sweep $R_{xx}$ curves, and the weak oscillations in
$R_{xy}$. The up-sweep curve has been offset to improve
clarity.}\label{1}
\end{center}
\end{figure}

A problem of interest is to identify the $B$ intervals that
exhibit vanishing resistance at a fixed $f$, and to connect them
in a simple way to $B_{f}$ because the relation between the two
scales $(hf, j\hbar \omega_{c})$, which is set by $B$, is likely
to be essential in the underlying physics. In earlier work, we had
reported that resistance minima occur about $B = [4/(4j+1)] B_{f}$
and resistance maxima transpire about $B = [4/(4j+3)] B_{f}$, with
$j$=1,2,3... \cite{1} This "minima about $B = [4/(4j+1)] B_{f}$"
result might also be loosely reformulated as a "1/4 cycle phase
shift" with respect to the $hf = j\hbar \omega_{c}$
condition.\cite{15}

Our "minima about $B = [4/(4j+1) B_{f}]$" result seems
incongruent, however, with a report which identified radiation
induced resistance minima with $\omega/\omega_{c} = j + 1/2$,
i.e., $B_{f}/B = j + 1/2$, and resistance maxima with
$\omega/\omega_{c} = j$, i.e., $B_{f}/B = j$, where $\omega = 2\pi
\emph{f}$.\cite{14} Briefly, that study of low mobility specimens
exhibited $T$-insensitive $R_{xx}$ oscillations, a large
non-vanishing resistance at the minima even at the lowest $T$, and
the absence of a phase shift.\cite{14} A subsequent study of high
mobility specimens identified resistance maxima with
$\omega/\omega_{c} = j$, as it placed resistance minima on the
high magnetic field side of $\omega/\omega_{c} = j + 1/2$.\cite{2}
\begin{figure}
\begin{center}
\includegraphics*[scale = 0.25,angle=0,keepaspectratio=true,width=3.25in]{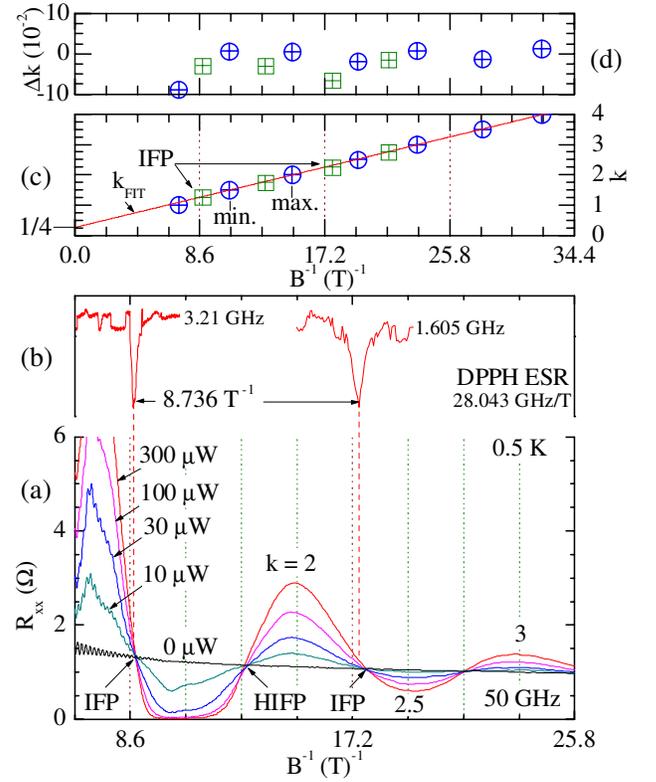}
\caption{(a) $R_{xx}$ is shown vs $B^{-1}$ for several radiation
intensities at 50 GHz. (b) ESR of DPPH for $f_{0} = 3.21$ GHz and
$f_{0}/2 = 1.605$ GHz. Here, $f_{0}$ marks the first Integral
Fixed Point (IFP) in (a). The period of the $R_{xx}$ oscillations,
8.6 $T^{-1}$, appears to be slightly smaller than the spacing of
the DPPH-ESR lines. (c) A half-cycle plot of the extrema in (a). A
linear fit suggests a 1/4 cycle phase shift. Fixed points,
inserted as open squares, were not used for the fit. (d) The
residue in $k$, i.e., $\Delta k = k - k_{FIT}$, extracted from
(c).} \label{2}
\end{center}
\end{figure}

A recent reformulation\cite{16} asserts that the phase of the
$R_{xx}$ oscillations in ref.\cite{2} are consistent with a 1/4
cycle phase shift,\cite{1} but only at low-$B$, i.e., $j \geq 4$.
Although this reinterpretation has helped to bring overlap between
theory,\cite{4,6,8} and recent experiment,\cite{1,2,12,13,16}
there continues to be a phase-variance in the regime of low $j$,
$j \leq 4$, between our initial report announcing these novel
zero-resistance states [ZRS], and the report by Zudov et al.,
confirming the same.\cite{1,2}
\begin{figure}
\begin{center}
\includegraphics*[scale = 0.25,angle=0,keepaspectratio=true,width=3.25in]{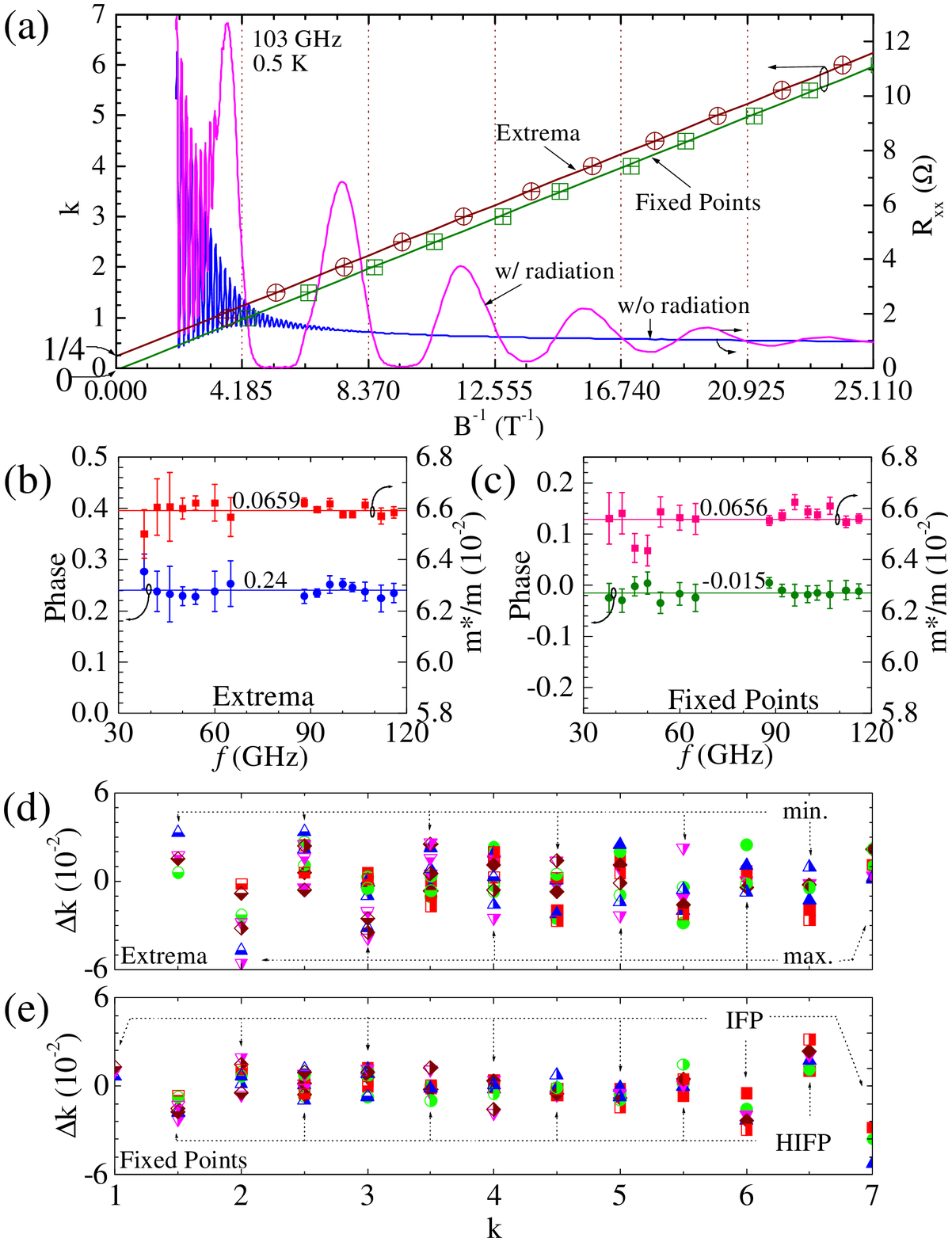}
\caption{(a)(right ordinate): $R_{xx}$ is plotted vs. $B^{-1}$
with (w/) and without (w/o) radiation at 103 GHz. (left ordinate)
Half-cycle plots of the extrema and the fixed points in the
radiation induced resistance oscillations. The slope of the linear
fit is proportional to $m^{*}/m$, while the ordinate intercept
measures the phase. (b): The phase and the effective mass ratio,
$m^{*}/m$, obtained from linear fits of half cycle plots of
extrema [see (a)] as a function of the frequency. (c): The phase
and $m^{*}/m$ obtained from linear fits of half cycle plots of
fixed points [see (a)] vs. $f$. (d): The residue in the extrema
labels, $\Delta k = k - k_{FIT}$, is shown as a function of $k$,
the assigned label, for $f$ shown in (b). (e): For the fixed
points, $\Delta k$  is shown vs. $k$, for $f$ in (c).} \label{3}
\end{center}
\end{figure}

Phase characterization in this context requires an accurate
measurement of $B$ since the observed oscillations are referenced
to $f$, which sets $B_{f}$. In the laboratory, the transport
set-up typically includes a high-field superconducting magnet, and
$B$ is often determined from the magnet coil current, $I$, via a
calibration constant. At low-$B$, $B \leq 0.5 T$, this approach to
$B$-measurement can be unreliable due to nonlinearity and
hysteresis in the $B - I$ characteristics of the field-coil, and
$B$ offsets due to trapped flux and/or current-offsets in the
magnet power supply. An in-situ Hall probe can be applied to
correct this problem. However, even this approach is prone to
field measurement errors arising from misalignment-voltages in the
Hall sensor or changes in the Hall constant.\cite{17} To reduce
such uncertainties, we have carried out microwave transport
studies of the 2DES with in-situ $B$-field calibration by Electron
Spin Resonance (ESR) of DiPhenyl-Picryl-Hydrazal (DPPH). The
results demonstrate a $f$-independent 1/4 cycle phase shift in the
radiation-induced oscillatory-magnetoresistance, and they suggest
a small reduction in the mass ratio characterizing this phenomena,
with respect to the standard value for $m^{*}/m$, $m^{*}/m =
0.067$.\cite{18}

Experimental details have been reported elsewhere.\cite{1,12,13}
Here, for $f \leq 50$ GHz, the samples were photo-excited with a
digital-frequency-readout microwave generator, with a 1
\textit{ppm } uncertainty in $f$ in the CW mode.\cite{19} At
higher frequencies, mechanical $f$-meters with an uncertainty of
0.4$\%$ above 100 GHz were used in conjunction with calibration
tables to determine $f$.\cite{20} Thus, $f$ is thought to be
better characterized below 50 GHz. ESR of DPPH was performed in
the immediate vicinity of the 2DES at $f \leq 10$ GHz using a
second digital-frequency-readout microwave source.\cite{21} The
g-factor for the unpaired electron undergoing ESR on DPPH is
2.0036, which corresponds to 28.043 GHz/T.\cite{22}

\begin{figure}
\begin{center}
\includegraphics*[scale = 0.25,angle=0,keepaspectratio=true,width=2.5in]{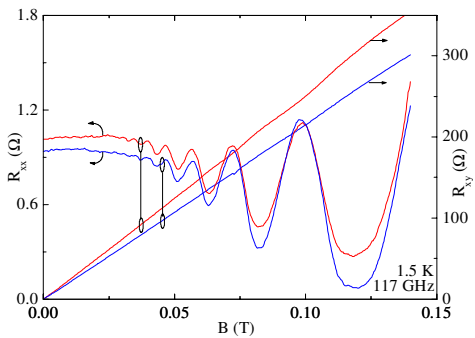}
\caption{ The periodicity of radiation induced $R_{xx}$
oscillations appears insensitive to the electron density, $n$.
Here, increasing the slope of $R_{xy}$ vs. $B$ corresponds to
decreasing $n$.} \label{1}
\end{center}
\end{figure}
Fig. 1(a) shows the magnetoresistance $R_{xx}$ measured in a
GaAs/AlGaAs device with microwave excitation at 30 GHz. Also shown
is the ESR of DPPH at 2.01 GHz, which marks $B_{f} = 2\pi f
m^{*}/e$ for $f$ = 30 GHz and $m^{*}/m$ = 0.067. The data of Fig.
1 (a) indicate that the ESR field marker coincides neither with a
minimum nor with a maximum in the $R_{xx}$ signal.\cite{1} Fig.
1(a) also indicates a strong resistance minimum below $B_{f}$,
about $4/5 B_{f}$, which is the $j = 1$ member of $B = (4/[4j+1])
B_{f}$. These data at 30 GHz also exhibit a deep minimum between
$B_{f}$ and $2B_{f}$, which is identified here as the possible
principal minimum for a two-photon process with a fundamental
field $2 B_{f}$, expected to form about $2(4/5 B_{f}) = 8/5
B_{f}$. The data suggest a convergence to $8/5 B_{f}$, as the
minimum becomes deeper with increasing microwave power.

Fig. 1(b) shows $R_{xx}$ under photoexcitation at 50 GHz. This
panel includes DPPH-ESR field markers which identify $B_{f}$, $4/5
B_{f}$, $B_{f}/2$ and $B_{f}/2.5$, for $f$ = 50 GHz. These data
show that, at $\pm B_{f}$, there is once again neither a maximum
nor a minimum in the $R_{xx}$, as in Fig. 1(a). In addition, a
zero-resistance state forms about the $\pm 4/5 B_{f}$ field
marker, and there is neither a maximum nor a minimum in $R_{xx}$
near the $\pm B_{f}/2$ marker.\cite{1} Thus, the data of Fig. 1
indicate that the $R_{xx}$ maxima do not coincide with
$B_{f}/j$.\cite{1,2,14,16} They also confirm the absence of
intrinsic sweep-direction dependent hysteretic effects.

Fig. 2(a) exhibits $R_{xx}$ vs. $B^{-1}$ at 50 GHz for several
radiation intensities. This plot shows regular-in-$B^{-1}$
$R_{xx}$ oscillations with a number of "fixed points" where
$R_{xx}$ is independent of the microwave power. According to the
theory,\cite{8} the conductance under radiation should equal the
dark conductance at $B_{f}/B = j$. We denote these as Integral
Fixed Points (IFP). Experiment also indicates Half-Integral Fixed
Points (HIFP) in the vicinity of $B_{f}/B = j + 1/2$.\cite{12} The
$B$-measurement of IFP's by ESR of DPPH is shown in Fig. 2(b).
Here, the ESR of DPPH at $f_{0}$ = 3.21 GHz marks the first IFP in
the $R_{xx}$ data of Fig. 2(a). Fig. 2 (a) and (b) indicate that,
when $f_{0}$ is selected to mark the first IFP, $f_{0}$/2 marks a
$B^{-1}$ that is slightly below the second IFP, suggesting a small
correction. This correction is attributed here to the observed
asymmetry between the resistance increase- and the resistance
decrease- with respect to the dark value. That is, as the $R_{xx}$
increase under excitation can easily exceed, in magnitude, the
$R_{xx}$ decrease, $R_{xx}$ peaks tend to span a wider $B^{-1}$
interval than the $R_{xx}$ valleys. This tends to relocate the IFP
to a slightly higher $B^{-1}$ value, than where the condition
$2\pi f /\omega_{c} = j$ actually holds true.

Another method for determining the period and the phase of the
oscillations involves the application of a half-cycle plot, shown
in Fig 2(c). Typically, the procedure assumes a cosine waveform.
Thus, the $R_{xx}$ maxima and minima are labelled by integers and
half-integers, respectively. These labels, denoted here by the
index $k$, are plotted vs. extremal values of $B^{-1}$, as in Fig.
2(c).  A linear fit, $k_{FIT}$, then serves to determine the
slope, $dk_{FIT}/dB^{-1}$ = $B_{f}$, and the ordinate intercept.
Here, in Fig. 2(c), the ordinate intercept at 1/4 demonstrates a
1/4 cycle phase shift of $R_{xx}$ with respect to the assumed
(cosine) waveform.\cite{13} The residual difference, $\Delta k$,
between $k$ and $k_{FIT}$ for the extrema and fixed points, has
been shown in Fig. 2(d). Fig. 2(d) confirms a small phase
distortion, discussed above, at the lowest fixed points. It also
confirms that the $k$ = 1 $R_{xx}$ maximum, which tends to be
upshifted from $B_{f}/B$ = 3/4, is a special case,\cite{12} see
also Fig. 3(a).

There are two parameters in the half-cycle analysis: the intercept
and the slope. Yet, the root-parameter in the slope is $m^{*}$
since $B_{f} = 2\pi f m^{*}/e$, and $f$ can be measured. The
question then arises whether the measured $m^{*}$ agrees with
expectations.

In order to obtain a representative sampling of the phase and
$m^{*}/m$, we have carried out a half-cycle analysis of a data set
obtained over a range $f$. For this work, both the extrema and the
fixed points were subjected to a separate half-cycle analysis at
each frequency, as illustrated in Fig. 3(a) for $f$ = 103 GHz.
Since the analysis of the extrema has been discussed above, we
note only that an analogous half-cycle analysis of the fixed
points involves associating IFP's with integers and HIFP with
half-integers, and plotting the resulting index $k$ vs. $B^{-1}$.

A linear least squares fit served to determine the slope and the
ordinate intercept of the $k$ vs. $B^{-1}$ plots, as in Fig. 3(a).
The characteristic field, $B_{f}$, and $m^{*}/m$ were extracted
using $B_{f}$ = $dk_{FIT}/dB^{-1}$, and $m^{*}/m$ = $e B_{f}/(2\pi
f m)$, where m is the electron mass. The resulting phase and
$m^{*}/m$ have been shown in Fig. 3(b) and Fig. 3(c). Fig. 3(b)
shows that the average phase for the extrema is 0.24 ($\pm$
0.015), which is 1/4 within uncertainty. On the other hand, Fig.
3(c) shows that the average phase for the fixed points is -0.015
($\pm$ 0.012), which is approximately zero. Thus, the phase
related results are broadly consistent with resistance minima at
$B_{f}/B = j + 1/4$, resistance maxima at $B_{f}/B = j + 3/4$,
IFP's at $B_{f}/B$ = $j$, and HIFP's at $B_{f}/B$ = $j + 1/2$, for
j = 1,2,3...\cite{12} The residual difference, $\Delta k$ = $k -
k_{FIT}$, for the extrema and fixed points have been shown in Fig.
3(d) and Fig 3(e), respectively. Here, the scatter $\Delta k$
spans a smaller band in Fig. 3(e) than in Fig. 3(d). Doublet
formation on the $k = 2$ maximum at large $f$ or low $T$, akin to
"spin splitting," has been found to introduce scatter in Fig.
3(d).

So far as the mass ratios are concerned, Fig. 3(b) suggests
$m^{*}/m$ = 0.0659 ($\pm$ 0.0004), and Fig. 3(c) indicates that
$m^{*}/m$ = 0.0656 ($\pm$ 0.0006). These mass ratios are roughly
2$\%$ below the standard value, $m^{*}/m$ = 0.067,\cite{18} and
1$\%$ below our previously reported result, $m^{*}/m$ =
0.0663.\cite{1} Here, the difference between the average mass of
Fig. 3 (b) and (c), and the standard value, appears to exceed the
resolution limit of this experiment.

Mass corrections can originate from, for example, the polaron
effect,\cite{23} where an electron polarizes the lattice and
creates phonons, and the dressing of the electron by the phonon
cloud leads to a mass change, usually a mass increase. An observed
mass reduction could be related to some such renormalization
effect, if the associated effect is somehow switched off in this
context, so that the measurement comes to reflect a bare mass,
which then looks like a mass reduction. A theoretical study
appears necessary to examine such a possibility. Band
non-parabolicity can lead to a dependence of $m^{*}/m$ on the
electron density, $n$, or the Fermi energy. Fig. 4 shows, however,
that the $R_{xx}$ oscillation period remains unchanged as $n$ is
varied by $\approx$ 15 $\%$ using the persistent photo-effect.

In summary, we have demonstrated a $f$-independent 1/4 cycle phase
shift of the extrema in the radiation induced
oscillatory-magnetoresistance with respect to the
$\omega/\omega_{c} = j$ or $B_{f}/B = j$ condition for $j \geq
$1.\cite{4,6,8} We also find that $m^{*}/m$ is 1 $\%$ smaller than
previously reported,\cite{1} and about 2 $\%$ smaller than the
standard value.\cite{18}

\vspace{0cm}


\begin{thebibliography}{24}
\bibitem{15} Minima about $B = [4/(4j-1)] B_{f}$ would also constitute a
1/4 cycle phase shift, although in the other direction.
\end{thebibliography}
\end{document}